\documentclass{elsarticle}

\usepackage{amssymb}
\usepackage{amsfonts}

\usepackage{amsthm}
\usepackage{amsmath}
\usepackage{natbib}

\theoremstyle{plain}
\newtheorem{theorem}{Theorem}[section]

\newtheorem{lemma}[theorem]{Lemma}

\numberwithin{equation}{section}

\newtheorem{thm}{Theorem}[section]

\newcommand{\be}{\begin{eqnarray}}
\newcommand{\ee}{\end{eqnarray}}
\newcommand{\ba}{\begin{array}}
\newcommand{\ea}{\end{array}}
\newcommand{\bc}{\begin{center}}
\newcommand{\ec}{\end{center}}

\newcommand{\bew}{\begin{eqnarray*}}
\newcommand{\eew}{\end{eqnarray*}}
\newcommand{\f}[2]{\frac{#1}{#2}}

\newcommand{\la}[1]{\label{eq:#1}}

\newcommand{\rf}[1]{(\protect\ref{eq:#1})}

\newcommand{\ct}[1]{\cite{#1}}

\newcounter{letter}
\newcounter{hold}
\newcommand{\blett}{\setcounter{hold}{\value{equation}}\addtocounter{hold}{1}}
\newcommand{\lett}{\setcounter{equation}{\value{hold}}\addtocounter{letter}{1}}
\newcommand{\elett}{\setcounter{letter}{0}}

\theoremstyle{definition}
\newtheorem{dfn}[thm]{Definition}

\journal{Journal of mathematical analysis and applications}

\begin{document}

\begin{frontmatter}



\title{The relations of Choquet Integral and G-Expectation }


\author{Ju Hong Kim}

\address{Department of Mathematics, Sungshin Women's University, Seoul 02844, Republic of Korea} 


\begin{abstract}
In  incomplete financial markets, there exists a set of equivalent martingale  measures (or risk-neutral probabilities) in an arbitrage-free  pricing of the contingent claims. Minimax expectation is closely related to the $g$-expectation which is 
 the solution of a certain stochastic differential equation.
 We show that Choquet expectation and minimax expectation are equal in pricing European type options, whose payoff is a monotone function of the terminal stock price $S_T$.

\end{abstract}


\end{frontmatter}


\section{Introduction} 
Nonlinear expectations such as Choquet expectation, minimax expectation and  $g$-expectation are applied to many areas like statistics, economics and finance. 
 Choquet expectation~\ct{Choquet53} has a difficulty in defining a  conditional expectation.
Wang~\ct{HongxiaWang15} introduces the concept of conditional Choquet expectation which is the conditional expectation with respect to a submodular capacity.

Choquet expectation~\ct{ChenChenDavison05, FollmerSchied04} is equivalent to the convex(or coherent) risk measure if given capacity is submodular.
 $G$-expectation (see papers\ct{CoquetHuMeminPeng02, RosazzaGianin06, HeHuChen09, Jiang06, LongJiang09,KarouiPengQuenez97, PardouxPeng90} for the related topics)  is the solution of the following nonlinear backward stochastic differential equation(BSDE),
\be 
   y_t = \xi + \int_t^T g(s, y_s , z_s ) ds - \int_t^T z_s dB_s, \quad 0 \leq t \leq T. \la{BSDE_INTRO}
\ee

$G$-expectation very much depends on the generator $g$ in the BSDE~\rf{BSDE_INTRO}.
If $g$ is sublinear with respect to $z$, then  $g$-expectation is represented as
\bew
  y_0 = \sup_{Q \in \mathcal P } E_Q [ \xi] \quad \quad \forall \xi \in L^2 (\Omega, \mathcal F_T , P)
\eew where $y_t$ is the solution of the BSDE~\rf{BSDE_INTRO}, $E_Q$ represents the expectation with respect to $Q$  and $\mathcal P$ is a set of risk-neutral probability measures.
Minimax expectation~\ct{ChenKulperger06} is the  expectation taken supremum or infimum over a set of probability measures.
Minimax expectation is very much related to $g$-expectation.

In this paper, we will show that Choquet and minimax expectations are equal in pricing European type options,
whose payoff is a monotone function of the terminal stock price $S_T$.
First, it is shown that the Choquet and minimax expectations are equal on the space of real-valued, bounded, $\mathcal F_T$-measurable functions,  $\mathcal B ( \Omega, \mathcal F_T,P)$. 
Second,  the function space of $\mathcal B ( \Omega, \mathcal F_T, P )$ is extended to 
a monotone subset of $L^2 (\Omega, \mathcal F_T , P)$.

\section{$G$-expectation and Choquet expectation}
In this section, we define the upper and the lower Choquet expectations,  and also find 
 the specific solution of the BSDE~\rf{BSDE_INTRO} 
when the generator  $g$ is sublinear with respect to $z$.
Minimax pricing rules are closely related to $g$-expectation, the solution of the BSDE~\rf{BSDE_INTRO}.

 Let $(\Omega, \mathcal F, P)$ be a given completed probability space.
Let $(\Omega, (\mathcal F_t)_{t\in[0,T]}, P)$ be the given filtered probability space.
The filtration $\mathcal F_t = \sigma\{ B_s : s \leq t\}$ is generated by 
$(B_t)_{t\in[0,T]}$, a one-dimensional standard Brownian motion.
Let $\mathcal B ( \Omega, \mathcal F_T, P )$ be the space of real-valued, bounded, $\mathcal F_T$-measurable functions, and
 let $V: \mathcal B ( \Omega,\mathcal F_T, P ) \rightarrow \mathbb R$ be a functional.

\begin{dfn}
A set function $c: \mathcal F_T \rightarrow [0,1]$ is called {\em  monotone} if 
\bew
 c(A) \leq c(B)  \quad \mbox{ for } A \subset B, \mbox{ and } A, B \in \mathcal F_T
\eew
and {\em normalized} if 
\bew
    c(\emptyset)=0\,\, \mbox{  and  } \,\, c(\Omega) =1.
\eew
The monotone and normalized set function is called a {\em capacity}.
A monotone set function is called {\em submodular} or {\em $2$-alternating} if 
\bew
 c(A\cup B) +c(A\cap B) \leq c(A) +c(B)  \quad A, B \in \mathcal F_T.
\eew
\end{dfn}
The risk of an asset position $X+Y$ will be lower than the sum of each risk,
because of the diversification effects.
The property of comonotonicity is that if there is no way for $X$ to serve as a hedge for $Y$, then it is  simply  adding up the risks.

Two real functions $X, Y \in \mathcal B (\Omega, \mathcal F_T, P ) $
are called {\em comonotonic} if
\bew
[X(\omega_1) -X(\omega_2)][Y(\omega_1) -Y(\omega_2)] \geq 0, \quad\omega_1, \omega_2 \in\Omega.
\eew

The functional $V$ is said to be {\em comonotonic additive} if 
\bew
   X, Y \mbox{ are comonotonic}  \Longrightarrow V(X+Y) = V(X) + V(Y).
\eew
\begin{dfn}
Let $c: \mathcal F_T \rightarrow [0,1]$ be a capacity.
The {\em Choquet expectation}   with respect to $c$
is defined as
\bew
\int_\Omega X\,dc := \int_{-\infty}^0 (c(X>x)-1)\, dx + \int_0^\infty c(X>x)\, dx,  \quad X\in L^2 (\Omega, \mathcal F_T , P).
\eew
\end{dfn}

The following  theorem of Schmeidler~\ct{Schmeidler86} tells us that there exists a capacity that the normalized, monotone, and
comonotonic additive functional is equal to Choquet expectation on $\mathcal B ( \Omega,\mathcal F_T, P)$.

\begin{thm}[\ct{Schmeidler86}]
Let $V$ be a functional from $ \mathcal B ( \Omega,\mathcal F_T, P) $ to $\mathbb R$. The following 
statements are equivalent.\label{Schme}
\begin{enumerate}
  \item $V$ is normalized, monotone, and comonotonic additive.
  \item There exists a unique capacity $c: \mathcal F_T \rightarrow [0,1]$ such that         
\be
 V(X) = \int_{-\infty}^0 (c(X>x)-1)\, dx + \int_0^\infty c(X>x)\, dx \la{VofXnChoquet}
\quad\forall X \in \mathcal B ( \Omega,\mathcal F_T, P). \,\,\la{V(X)}
\ee
\end{enumerate}
\end{thm}

Let  $g: \Omega \times [0,T] \times \mathbb R \times \mathbb R^n \rightarrow \mathbb R$ be  a function
that $(y,z)\mapsto g(t,y,z)$ is measurable for each $(y,z)\in \mathbb R \times \mathbb R^n$ and 
satisfy  the following conditions
\blett
\be\la{gcondition}
       &&|g(t,y,z)-g(t, \bar y, \bar z)| \leq K (|y- \bar y| + |z- \bar z|)\lett\la{gconA1}\\
          &&\qquad \forall t\in[0,T], \forall (y,z),(\bar y, \bar z)\in \mathbb R \times \mathbb R^n, \mbox{ for some } K >0,  \nonumber \\
     && \int_0^T |g(t,0,0)|^2 \, dt <\infty, \lett\la{gconA2}\\
&& g(t,y,0) =0 \mbox{ for each } (t,y) \in [0,T] \times \mathbb R. \lett\la{gconA3}
\ee
\elett 

The space $L^2(\Omega, \mathcal F_T, P)$ is defined as
\bew
    L^2(\Omega, \mathcal F_T, P):=\{ \xi \,\,| \,\xi \mbox{ is $\mathcal F_T$-measurable random variable and  }
        E[|\xi|^2 ] <\infty\}.
 \eew

\begin{thm} [\ct{Peng97}]
For every terminal condition $\xi\in L^2(\mathcal F_T ):=L^2(\Omega,\mathcal F_T, P)$ the following backward stochastic differential equation 
\blett
\be\la{BSDE}
    -dy_t &=& g(t, y_t, z_t) \, dt - z_t d B_t, \quad 0\leq t \leq T\lett, \\
       y_T &=& \xi \lett
\ee
\elett 
has a unique solution 
\bew
(y_t, z_t)_{t\in[0,T]} \in L_{\mathcal F}^2([0, T]; \mathbb R)\times L_{\mathcal F}^2 ( [0,T]; \mathbb R^n).
\eew
\end{thm}

\begin{dfn}
For each $\xi\in L^2(\mathcal F_T )$ and for each $t\in [0,T]$ {$g-$expectation} of $\xi$ and 
the conditional $g-$expectation of $\xi$ under $\mathcal F_t$ is respectively defined by 
\bew
  \mathcal E_g [\xi ] := y_0, \quad \mathcal E_g [\xi|\mathcal F_t ] := y_t,
\eew where $y_t$ is the solution of the BSDE~\rf{BSDE}.
\end{dfn}

Let $\{ S_t \}$ be the stock price evolving as a stochastic differential equation
\bew
 \f{ dS_t}{ S_t} = \mu_t dt +\sigma_t  dB_t
\eew where $\{\mu_t\}$ is a market return rate, and $\{\sigma_t\}$ is a market volatility.

In a Black-Scholes world, there exists a unique risk-neutral probability measure $Q$ defined as 
\bew
\f{dQ}{dP} = e^{ - \f{1}{2} \int_0^T \left ( \f{\mu_s -r}{\sigma_s} \right )^2 ds 
+\int_0^T \left ( \f{\mu_s -r}{\sigma_s} \right ) dB_s},
\eew where $r$ is a riskless interest rate.
In a real world, the parameters $\mu_t$ and $\sigma_t$ are not known exactly. 
We assume that $\mu_t$ belong to some interval, i.e. $\mu_t \in[ r-k \sigma_t , r+k \sigma_t]$ for a constant $k >0$.
Then the risk-neutral probability measures belong to 
\bew
\mathcal P = \left \{ Q^\nu \,\, : \,\,
\f{dQ^\nu}{dP} = e^{ - \f{1}{2} \int_0^T  |\nu_s|^2 ds 
+\int_0^T \nu_s dB_s}, \,\, \sup\limits_{t\in[0,T]} |\nu_t|\leq k 
             \right \}
\eew where $\nu_t := (\mu_t -r)/\sigma_t$.
There are two pricing methods of a contingent claim $\xi$, i.e. minimax pricing rules which are
\bew 
\underline{\mathcal E} [\xi]:= \inf\limits_{Q\in \mathcal P} E_Q[\xi], \quad \bar{\mathcal E} [\xi]:= \sup\limits_{Q\in \mathcal P} E_Q[\xi].
\eew

Let $\xi \in L^2(\Omega, \mathcal F_T, P)$.
The conditional $g$-expectations $\bar{\mathcal E} [\xi|\mathcal F_t ]$ and 
 $\underline{\mathcal E} [\xi|\mathcal F_t ]$ are   given as
\be
\bar{\mathcal E} [\xi|\mathcal F_t ]=\mbox{ess}\sup\limits_{Q\in \mathcal P} E_Q[\xi| \mathcal F_t], \qquad 
\underline{\mathcal E} [\xi|\mathcal F_t ] = \mbox{ess} \inf\limits_{Q\in \mathcal P} E_Q[\xi| \mathcal F_t],\la{minimax:eqn}
\ee which are the solutions of BSDE~\rf{BSDE} when the generators are $g(t, y_t, z_t)=k|z_t|$ and 
$g(t, y_t, z_t)=-k|z_t|$ respectively. The equations~\rf{minimax:eqn}  will be proved in Lemma~\ref{lemma21}.

It is clear that 
\bew
\bar{\mathcal E} [\xi|\mathcal F_0 ]= \bar{\mathcal E} [\xi]:=\sup\limits_{Q\in \mathcal P} E_Q[\xi], \qquad
\underline{\mathcal E} [\xi|\mathcal F_0 ]= \underline{\mathcal E} [\xi]:= \inf\limits_{Q\in \mathcal P} E_Q[\xi].
\eew

The upper and the lower Choquet integrals(or expectations) are respectively defined as
\bew 
&&\bar V(\xi) := \int_{-\infty}^0 (\bar c(\xi>x)-1)\, dx + \int_0^\infty \bar c(\xi>x)\, dx,\\
&&\underline V(\xi) := \int_{-\infty}^0 (\underline c(\xi>x)-1)\, dx + \int_0^\infty \underline c(\xi>x)\, dx, 
\eew where $\bar c$ and $\underline c$ are defined as 
\bew
 \bar c(A) =\sup\limits_{Q\in \mathcal P}Q(A) \quad \mbox{and}\quad 
\underline c(A) =\inf\limits_{Q\in \mathcal P}Q(A)\quad \mbox{ for } A \in \mathcal F_T.
\eew
We will use the notation of $\bar V(\xi) := \int \xi \, d {\bar c}$ and $\underline V(\xi) := \int \xi \, d {\underline c}$, or sometimes integration notation just for the  convenience of proof.

It can be easily seen that
\bew
 \underline V(\xi)\leq \underline{\mathcal E} [\xi] \leq \bar{\mathcal E} [\xi] \leq \bar V(\xi).
\eew

In the complete market where  $\mathcal P$ has a single element, we can see that
\bew
\underline V(\xi)= \underline{\mathcal E} [\xi] =  \bar{\mathcal E} [\xi] = \bar V(\xi).
\eew

\begin{thm}[\ct{ChenChenDavison05}]\label{Thm26}
Suppose that $g$ satisfies the condition \rf{gconA1}-\rf{gconA3}.  Then there exists a Choquet integral whose restriction 
to $L^2(\Omega, \mathcal F_T, P)$ is equal to a $g$-expectation if and only if $g$ does not depend on $y$ and is linear in $z$, that is, 
there exists a continuous function $\nu_t$ such that
\bew
g(y, z, t) = \nu_t z.
\eew
\end{thm}
The  Theorem~\ref{Thm26} implies that the generator $g$ in \rf{BSDE} should be linear function for both Choquet integral and $g$-expectation to be equal. We will show that $\mathcal {\bar E} [\xi|\mathcal F_t ]$ and $\mathcal {\underline E} [\xi|\mathcal F_t ]$
are the solutions of the BSDEs~\rf{essSupInf} in the following Lemma~\ref{lemma21}.

\begin{lemma}
\label{lemma21}
For $\xi \in L^2 (\Omega, \mathcal F_T , P)$, 
let $(Y_t, z_t)$ and  $(y_t, z_t)$ be the unique solution of the following BSDEs
\blett
\be \la{essSupInf}
   &&Y_t =\xi + \int_t^T k |z_s| \, ds - \int_0^T z_s \, dB_s,\quad  t\in[0,T],\lett\la{essSupYsubt}\\   
   &&y_t =\xi - \int_t^T k |z_s| \, ds - \int_0^T z_s \, dB_s,\quad  t\in[0,T]\lett\la{essInfysubt}
\ee respectively.
\elett
Then $Y_t$ and   $y_t$ 
are respectively represented as 
\blett
\be
&&Y_t =\mbox{\em ess} \sup\limits_{Q\in \mathcal P} E_Q[\xi| \mathcal F_t]
=\mathcal {\bar E} [\xi|\mathcal F_t ],\lett\la{BigYsubt}\\
&& y_t =\mbox{\em ess} \inf\limits_{Q\in \mathcal P} E_Q[\xi| \mathcal F_t]
=\mathcal {\underline E} [\xi|\mathcal F_t ].\lett\la{Smallysubt}
\ee
\elett
\end{lemma}

\begin{proof}
First, we show  \rf{BigYsubt}.
Let $\nu_t = k\, \mbox{sgn}(z_t)$. Then $\sup\limits_{t \in[0,T] } |\nu_t|\leq k$.
If we define $z_t^\nu $ as
\bew
z_t^\nu
=exp\left (- \frac{1}{2} \int_0^t | \nu_s|^2 ds+\int_0^t \nu_s dB_s \right ),\quad  0 \leq t \leq T,
\eew 
then $(z_t^\nu)_{0 \leq t \leq T}$ is a $P$-martingale since $dz_t^\nu /z_t^\nu =\nu_t \cdot dB_t$. 
Also $z_T^\nu$ is a $P$-density on $\mathcal F_T$ since $1= z_0^\nu = E[z_T^\nu]$.

Define an equivalent martingale probability measure $Q^\nu$ and a Brownian motion $\bar B_t$ as 
\bew
\f{dQ^\nu}{dP} = e^{ - \f{1}{2} \int_0^T  |\nu_s|^2 ds 
+\int_0^T \nu_s dB_s}, \qquad \bar B_t = B_t - \int_0^t \nu_s \, ds.
\eew
Then $Q^\nu \in \mathcal P$, and Girsanov's theorem implies that $\{ \bar B_t\}$ is a $Q^\nu$-Brownian motion.

The BSDE \rf{essSupYsubt} is expressed as 
\bew
 Y_t =\xi -\int_t^T z^\theta_s\, d \bar B_s.
\eew
So we get 
\be
Y_t =E_{Q^\nu} [\xi \,|\, \mathcal F_t] \leq \mbox{ess} \sup\limits_{Q\in \mathcal P} E_Q[\xi| \mathcal F_t].\la{YleqEssSup}
\ee

Let $\{ \theta_t\}$ be a adapted process satisfying 
\bew
    \sup_{t\in[0,T]} |\theta_t| \leq k.
\eew
Consider the following BSDE
\be
 Y^\theta_t =\xi + \int_t^T \theta_s z^\theta_s \, ds - \int_t^T z^\theta_s \, dB_s, \quad t \in[0,T].\la{YHatThetat}
\ee
Define an equivalent martingale probability measure $Q^\theta$ and a Brownian motion $\bar B^\theta_t$ as 
\bew
\f{dQ^\theta}{dP} = e^{ - \f{1}{2} \int_0^T  |\theta_s|^2 ds 
+\int_0^T \theta_s dB_s}, \qquad \bar B^\theta_t = B_t - \int_0^t \theta_s \, ds.
\eew
Then $Q^\theta \in \mathcal P$, and Girsanov's theorem implies that $\{ \bar B^\theta_t\}$ is a $Q^\theta$-Brownian motion.
The BSDE \rf{YHatThetat} is expressed as 
\bew
 Y^\theta_t =\xi -\int_t^T z_s\, d \bar B^\theta_s.
\eew
So we get 
\bew
Y^\theta_t =E_{Q^\theta} [\xi \,|\, \mathcal F_t]. 
\eew
Since $\theta_t z_t \leq k |z_t|$ for all $(z_t,t) \in \mathbb R \times [0,T]$, the Comparison Theorem applied to 
\rf{essSupYsubt} and 
\rf{YHatThetat}, implies
that 
\bew
E_{Q^\theta} [\xi \,|\, \mathcal F_t] = Y^\theta_t \leq Y_t \quad \forall  t \in [0,T]
\eew
Hence we obtain
\be
\mbox{ess} \sup\limits_{Q\in \mathcal P} E_Q[\xi| \mathcal F_t] \leq Y_t.\la{EssSupleqY}
\ee
The inequalities \rf{YleqEssSup} and \rf{EssSupleqY} implies that
\bew
\bar{\mathcal E}[\xi | \mathcal F_t]:= \mbox{ess} \sup\limits_{Q\in \mathcal P} E_Q[\xi| \mathcal F_t]
\eew is the solution of \rf{essSupYsubt}.

In the same fashion,  we can show that
\bew
\underline{\mathcal E}[\xi | \mathcal F_t]:= \mbox{ess} \inf\limits_{Q\in \mathcal P} E_Q[\xi| \mathcal F_t]
\eew is the solution of \rf{essInfysubt} by setting $\nu_t = -k\, \mbox{sgn}(z_t)$.

\end{proof}

\section{Choquet expectation and minimax expectation}
In this section, we show that Choquet expectation and minimax expectation are equal in pricing European type options, whose payoff is a monotone function of the terminal stock price $S_T$. 
We also prove that the minimax expectation attains a maximum or a minimum on the set of equivalent martingale probability measures 
which is weakly compact.

At the expiration date $T$, let the stock price $S_T \in L^2(\Omega, \mathcal F_T, P)$ be a unique solution of the following SDE
\bew 
 dS_t =\mu_t S_t \, dt + \sigma_t S_t dB_t, \quad t \in [0,T].
\eew
Let $\Phi$ be a monotone function such that $\Phi(S_T) \in L^2 (\Omega, \mathcal F_T, P)$.
Let $(Y_t, z_t)$ and  $(y_t, z_t)$ be the unique solution of the following BSDE
\bew
   &&Y_t =\Phi(S_T) + \int_t^T \mu_s |z_s| \, ds - \int_0^T z_s \, dB_s,\\   
   &&y_t =\Phi(S_T) - \int_t^T \mu_s |z_s| \, ds - \int_0^T z_s \, dB_s,
\eew respectively.

In Lemma~\ref{lemma21}, we have shown that 
\bew
Y_t =\bar{\mathcal E} [\Phi(S_T)|\mathcal F_t ], \qquad 
y_t =\underline{\mathcal E} [\Phi(S_T)|\mathcal F_t ].
\eew
For example, in the option pricing, the monotone functions $\Phi(x) = (x-K)^+$ or $\Phi(x) = (K-x)^+$ is the payoff function 
of European call or put option, respectively. Here $K$ is an exercise price of the option.
We want to show that 
\bew
 {\bar {\mathcal E}}[\Phi(S_T)] = {\bar V}[\Phi(S_T)], 
\quad {\underline {\mathcal E}}[\Phi(S_T)] = {\underline V}[\Phi(S_T)],
\eew where $\bar V$ and $\underline V$ are the upper and lower Choquet expectations,
respectively.

Since ${\bar{\mathcal E}}[\xi]$ is defined as
\bew
{\bar{\mathcal E}}[\xi] :=\sup\limits_{Q\in \mathcal P} E_Q[\xi  ],
\eew
 it is obvious that ${\bar{\mathcal E}}$ is normalized and monotone.


For each $i=1,2$, let the random variables $\xi_i's$ be comonotonic  functions. 

${\bar{\mathcal E}}$ is comonotonic additive since 
\bew
{\bar{\mathcal E}}[\xi_1 + \xi_2 ] = \sup\limits_{Q\in \mathcal P} E_Q[\xi_1 + \xi_2  ]=
\sup\limits_{Q\in \mathcal P} E_Q[\xi_1 ]+\sup\limits_{Q\in \mathcal P} E_Q[\xi_2 ]
={\bar{\mathcal E}}[\xi_1]+{\bar{\mathcal E}}[\xi_2].
\eew

So Theorem~\ref{Schme} says that there exists a unique capacity $c: \mathcal F_T \rightarrow [0,1]$
satisfying 
\be\la{EofXChoquet}
\quad {\bar{\mathcal E}}[X] = \int_{-\infty}^0 (c(X>x)-1)\, dx + \int_0^\infty c(X>x)\, dx \la{EofXChoquet}
\quad\forall X \in \mathcal B ( \Omega,\mathcal F_T, P).
\ee

If we take $X=I_A$ for $A\in \mathcal F_T$, then \rf{EofXChoquet} becomes
\be
  \bar {\mathcal E}[I_A] = \int_{-\infty}^0 (c(I_A>x)-1)\, dx + \int_0^\infty c(I_A>x)\, dx.
\ee
Thus we have 
     \be\la{Sup:capacity}
          c(A)=  \sup\limits_{Q\in \mathcal P} Q[A ]:= {\bar c}(A) .
     \ee
So we have $ c= \bar c$.

Therefore, the equation \rf{EofXChoquet} becomes
\bew
\quad {\bar{\mathcal E}}[X] = \int_{-\infty}^0 (\bar c(X>x)-1)\, dx + \int_0^\infty \bar c(X>x)\, dx:= \bar V(X)
\quad\forall X \in \mathcal B ( \Omega,\mathcal F_T, P).
\eew

From now on, we will show that the equation \rf{EofXChoquet} can be extended from $\mathcal B ( \Omega,\mathcal F_T, P)$ to
a set of the monotone functions  which is a subset of $L^2( \Omega,\mathcal F_T, P)$.

\begin{lemma} \label{cbarissubmodular}
The capacity $\bar c$ in \rf{Sup:capacity}
 is submodular.
\end{lemma}

\begin{proof}
 It's easily shown that $\bar c$ is monotone and normalized. 
Since  $I_{A\cup B}$ and $I_{A \cap B}$ are a pair of comonotone functions for all $A, B \in\mathcal F_T$,
the comonotonicity of $\bar {\mathcal E}$ implies
\bew
\bar c( A\cap B) + \bar c( A\cup B)&=& \bar {\mathcal E} [ I_{A \cap B}]+\bar {\mathcal E} [ I_{A\cup B}] 
= \bar {\mathcal E}[ I_{A\cap B} + I_{A \cup B}] \\
&=&\bar {\mathcal E} [ I_{A} + I_{ B}]\\
&\leq& \bar {\mathcal E} [ I_A]  + \bar {\mathcal E}[I_B]= \bar c (A ) +\bar c(B).
\eew

So the proof is done.
 \end{proof}

\begin{lemma}\label{L2continuity}
 $\bar{\mathcal  E}[\xi]:=\sup\limits_{Q\in \mathcal P} E_Q[\xi]$ is $L^2$-continuous for comonotonic functions $\xi \in L^2 (\Omega, \mathcal F_T, P)$.
\end{lemma}

\begin{proof}
Let $\xi_1$ and $\xi_2$ be comonotonic functions.
Since $\bar{\mathcal  E}$ is comonotonic additive,
\bew
 | \bar{\mathcal  E}[\xi_2] -\bar{\mathcal  E}[\xi_1] | &=& |\bar{\mathcal  E}[\xi_2 -\xi_1]| =
\Big|\sup\limits_{Q\in \mathcal P} E_Q[\xi_2-\xi_1]\Big | \\                
  &\leq&  \sup\limits_{Q\in \mathcal P} E_Q[|\xi_1 -\xi_2|] =\bar{\mathcal  E}[|\xi_2 -\xi_1|].
\eew
Now we'll show that $\bar{\mathcal  E}$ is $L^2$-bounded.
Let  an adapted process $\{\theta_t\}$ bounded by $k$ be such that
\bew
 \f{dQ^\theta}{dP} = e^{ -\f{1}{2} \int_0^T |\theta_s|^2 \, ds + \int_0^T \theta_s \, dB_s}.
\eew 


By the H\"{o}lder's inequality, we have
\bew
 E_{Q^\theta} (|\xi|) &=& E\left ( |\xi| \f{d Q^\theta}{dP} \right )\leq (E[|\xi|^2])^{\f{1}{2}} \left (E\left [\Big| \f{d Q^\theta}{dP}\Big |^2 \right]\right )^{\f{1}{2}}\\
&=&  (E[|\xi|^2])^{\f{1}{2}} \left (E\left [  e^{ -\f{1}{2} \int_0^T |2\theta_s|^2 \, ds + \int_0^T 2\theta_s \, dB_s +\int_0^T |\theta_s|^2\, ds} \right]\right )^{\f{1}{2}} \\
&=&  (E[|\xi|^2])^{\f{1}{2}} \left (e^{\int_0^T |\theta_s|^2\,ds}E\left [  e^{ -\f{1}{2} \int_0^T |2\theta_s|^2 \, ds + \int_0^T 2\theta_s \, dB_s } \right]\right )^{\f{1}{2}}\\
&\leq&  (E[|\xi|^2])^{\f{1}{2}} e^{ \f{1}{2} k^2 T}.
\eew
So we get 
\be
\bar{\mathcal  E}[\xi]=\sup\limits_{Q\in \mathcal P} E_Q[\xi] \leq (E[|\xi|^2])^{\f{1}{2}} e^{ \f{1}{2} k^2 T}.\la{MathcalEL2bnd}
\ee
Thus we have
\bew
 | \bar{\mathcal  E}[\xi_2] -\bar{\mathcal  E}[\xi_1] | \leq (E[|\xi_2 -\xi_1|^2])^{\f{1}{2}} e^{ \f{1}{2} k^2 T}.
\eew
Therefore, $\bar{\mathcal  E}$ is $L^2$-continuous for the comonotonic random variables.
\end{proof}

On $ L^2 (\Omega, \mathcal F_T, P)$, denote Choquet integral  as 
\bew
  \int_{\Omega} X \, dc : = \int_{-\infty}^0 (c(X > x)-1)\, dx + \int_0^\infty c(X>x)\, dx \quad \forall X \in  L^2 (\Omega, \mathcal F_T, P),
\eew just for the  convenience of proof.
\begin{thm}[\ct{CerdaMartinSilvestre11}]
\label{HolderInequality}
Let $X,Y$ be real-valued measurable functions defined on $\Omega$.
If a capacity $c$ is submodular and $1< p,q <\infty$ with $\f{1}{p}+ \f{1}{q} =1$, then
\bew
 \int_\Omega |XY| \, dc \leq \left (\int_\Omega  |X|^p \, dc\right )^\f{1}{p} \left (\int_\Omega  |Y|^q \, dc\right )^\f{1}{q}.
\eew
\end{thm}

The following is the main theorem.
\begin{thm}
Let $X\in  L^2 (\Omega, \mathcal F_T, P)$ be a monotone function. Then we have
\bew
\bar {\mathcal E}[X]  = \int_\Omega X\, d \bar c, \quad \underline {\mathcal E} [X]  = \int_\Omega X\, d \underline c.
\eew
\end{thm}

\begin{proof}
Since $\underline {\mathcal E} [X] =-\bar {\mathcal E}[-X]$, we only prove that $\bar {\mathcal E}[X]  = \int_\Omega X\, d \bar c$.
Let $X\in  L^2 (\Omega, \mathcal F_T, P)$ be a monotone  random variable.  
Let $f$ be a simple function. Let $\epsilon >0$ be given.
\be
\left | \bar {\mathcal E}[X] -  \int_\Omega X\, d \bar c \right | &\leq &| \bar {\mathcal E}[X] -\bar {\mathcal E}[f]| +\left |\bar {\mathcal E}[f] - \int_\Omega f\, d \bar c \right  |  \nonumber \\
&&+\left | \int_\Omega f\, d \bar c - \int_\Omega X\, d \bar c\right |.\la{Ineqality3}
\ee
Since simple functions are dense in $L^2 (\Omega, \mathcal F_T, P)$, there exists an increasing simple function $f\nearrow X$ satisfying both
\bew
      &&  \| X-f\|_{L^2} < e^{-\f{1}{2}k^2 T}\cdot\f{\epsilon }{3} \quad \mbox{ and }\\
      &&\left (\int_\Omega  |f-X|^2 \, d\bar c\right )^\f{1}{2} = 
\left (\int_0^\infty \bar c( |f-X|^2>x) \, dx\right )^\f{1}{2} <\f{\epsilon}{3}.
\eew
Since the $\bar{\mathcal E}$ is $L^2$-continuous for the comonotonic random variables $X$ and $f$ by Lemma~\ref{L2continuity}, the first term of the right hand side of \rf{Ineqality3} is less than $\epsilon/3$.
 The equation \rf{EofXChoquet} implies that
the second term of the right hand side of \rf{Ineqality3} is zero.

The capacity $\bar c$ is submodular by Lemma~\ref{cbarissubmodular} and so Theorem~\ref{HolderInequality} implies that 
the third term of the right hand side of \rf{Ineqality3} becomes
\bew
\left | \int_\Omega f\, d \bar c - \int_\Omega X\, d \bar c\right |\leq  \int_\Omega |f-X|\, d \bar c 
&\leq &
\left (\int_\Omega  |f-X|^2 \, d\bar c\right )^\f{1}{2} \left (\int_\Omega  1_\Omega^2 \, d \bar c\right )^\f{1}{2}\\
&\leq& \left (\int_\Omega  |f-X|^2 \, d\bar c\right )^\f{1}{2} < \f{\epsilon} {3}.
\eew
So we obtain 
\bew
\left | \bar {\mathcal E}[X] -  \int_\Omega X\, d \bar c \right | < \epsilon.
\eew
Therefore, the proof is done.
\end{proof}

We will show that there exists $Q\in \mathcal P$ such that the minimax expectation takes a maximum or minimum. 
\begin{lemma}
The set of densities 
\bew 
 \mathcal D:= \left \{ \f{dQ}{dP} \,\Big | \, Q \in \mathcal P \right \}
\eew
is weakly compact in $L^2(\Omega, \mathcal F_T, P)$.
\end{lemma}

\begin{proof}
As in the proof of Lemma~\ref{L2continuity}, we can prove
\bew
  E\left [ \left (  \f{d Q}{dP} \right )^2\right ]\leq e^{ \f{1}{2} k^2 T}.
\eew
So we have
$\f{dQ}{dP} \in L^2(\Omega, \mathcal F_T, P)$. Thus we have $\mathcal D \subset L^2(\Omega, \mathcal F_T, P)$ .

We want to show that $\mathcal D$ is weakly closed in $L^2(\Omega, \mathcal F_T, P)$.
Suppose that the sequence $(Z_n)$ in $\mathcal D$ converges weakly to $Z$.
I.e., 
\bew
   f(Z_n) \rightarrow f(Z) \mbox{ for all } f \in (L^2)^*, \mbox{ where }  (L^2)^* \mbox{ is the set of  continuous dual functionals of  } L^2.
\eew
We want to show $Z \in \mathcal D$.

For $X \in L^2(\Omega, \mathcal F, P)$, define the linear functional $J_X$  as 
\be\la{lin:fun}
  J_X (Z) := E[XZ]  \quad \forall Z \in \mathcal D.
\ee
By the H\"{o}lder's inequality, we have
\bew
 |J_X (Z)|\leq E[|XZ|]  \leq \left ( \int |X|^2 dP\right )^{1/2} \cdot \left ( \int |Z|^2 dP\right )^{1/2} < +\infty.
\eew

So $J_X$ is bounded and thus continuous on $L^2$.

By the assumption, we have
\bew
   J_X(Z_n) \rightarrow J_X(Z)  \mbox{ as } n \rightarrow \infty.
\eew
That is,
\bew
\lim_{n\rightarrow\infty} \int X dQ_n=\lim_{n\rightarrow\infty}E[XZ_n]= E[XZ]=\int X dQ.
\eew

Since $Z_n \in \mathcal D$, there exist $\theta_t^{(n)}$ and $Q^{\theta_t^{(n)}}\in \mathcal P$ satisfying
\bew
     Z_n = \f{ dQ^{\theta^{(n)}}} {dP} = \exp \left ({ -\f{1}{2} \int_0^T |\theta_s^{(n)}|^2 \, ds + \int_0^T \theta_s^{(n)} \, dB_s} \right).
\eew

Let $\lim_{n\rightarrow \infty}\theta_t^{(n)} = \theta_t$. Then we have 
\bew
 Z' = \lim_{n\rightarrow \infty} Z_n =  \exp\left ({ -\f{1}{2} \int_0^T |\theta_s|^2 \, ds + \int_0^T \theta_s \, dB_s}\right).
\eew
So we have 
\bew
  \int_0^T XZ dP = \int_0^T X Z' dP \quad  \forall  X \in L^2(\Omega, \mathcal F_T, P). 
\eew
Therefore, it becomes $ Z = Z' \quad a.e.$ and thus $Z \in \mathcal D$.
It is proven that $\mathcal D$ is a weakly compact set.
\end{proof}

\begin{thm}[James' Theorem]
A weakly closed subset $\mathcal D$ of a Banach space $L^2(\Omega, \mathcal F_T, P)$  is weakly compact if and only if each continuous linear functional on $L^2(\Omega, \mathcal F_T, P)$  attains a maximum or a minimum on $\mathcal D$.
\end{thm}

By James' Theorem, the linear functional $J_X$ as in \rf{lin:fun} attains 
a maximum on  $\mathcal D$.
That is, there exists $Q^* \in \mathcal P$ such that 
\bew
\sup_{Q\in \mathcal P}E_Q [ \xi] = E_{Q^*} [ \xi]   \quad \xi \in L^2(\Omega, \mathcal F_T, P) .
\eew

To specify  $Q^* \in \mathcal P$, we need Lemma~\ref{ChenChenDav} 
which gives the restriction to the generator $g$ of BSDE~\rf{SlnYsubt}, in addition to Theorem~\ref{Thm26}.
Let $\{ S_t\}$ be the solution of the following stochastic differential equation,
\be
   S_t = S_0 + \int_0^t \eta(t, S_t) dt + \int_0^t \sigma(t, S_t ) dB_t, \quad t \in[0, T],\la{SDE}
\ee where $\eta$, $\sigma : [0,T] \times \Re  \rightarrow \Re$ are continuous in $(t, S)$ and Lipschitz continuous in $S$.
\begin{lemma}[\ct{ChenChenDavison05}]\label{ChenChenDav} 
Let $\{ S_t\}$ be the solution of \rf{SDE}. Let $\Phi$ be the monotone function such that 
$\Phi(S_T) \in L^2(\Omega, \mathcal F_T , P)$.
Let $(y_t, z_t)$ be the solution of the following BSDE
\be
y_t = \Phi(S_T ) + \int_t^T \theta_s |z_s| -\int_t^T z_s dB_s.\la{SlnYsubt}
\ee
Then the followings hold,
\begin{enumerate}
 \item   $z_t \sigma(t,S_t ) \geq 0$, a.e. $t \in [0,T)$, \mbox{ if $\Phi$ is an increasing function}     
 \item  $z_t \sigma(t,S_t ) \leq 0$, a.e. $t \in [0,T)$, \mbox{ if $\Phi$ is a decreasing function}. 
\end{enumerate}
\end{lemma}

Suppose that $\Phi$ is an increasing function.  
Then for $|\theta_s | \leq k$, by Theorem~\ref{ChenChenDav},
the solution $(y_t, z_t)$ of \rf{SlnYsubt} becomes the unique solution of the form of BSDE
\be
y^{(\theta)}_t &=&\Phi(S_T) + \int_t^T \theta_s z^{(\theta)}_s \, ds - \int_t^T z^{(\theta)}_s \, dB_s\la{eqn1}\\
               &=& \Phi(S_T) -\int_t^T z^{(\theta)}_s \, d {\bar B}^\theta_s, \nonumber
\ee where ${\bar B}^\theta_t = B_t -\int_0^t \theta_s ds$.

Let  $(y^{(k)}_t, z^{(k)}_t)$  be the unique solution of the following BSDE
\be
  y^{(k)}_t =\Phi(S_T) + \int_t^T k |z^{(k)}_s| \, ds - \int_t^T z^{(k)}_s \, dB_s. \la{eqn2}
\ee

As we did at the end of Section $2$, 
we have $y^{(k)}_t \geq y^{(\theta)}_t$ for all  $t \in [0,T]$
by applying the Comparison Theorem for BSDEs to \rf{eqn1} and \rf{eqn2}.
Therefore, we get
\bew
   y^{(k)}_0=   E_{Q_k} [ \Phi(S_T)] \geq y^{(\theta)}_0=   E_{Q_\theta} [ \Phi(S_T)]
\eew where $Q_k$ and $Q_\theta$ are respectively defined as
\bew
\f{d Q_k}{dP} = e^{-\f{1}{2} \int_0^T k^2 ds + \int_0^T k dB_s}=e^{-\f{1}{2}k^2 T +kB_T},\quad
\f{d Q_\theta}{dP} = e^{-\f{1}{2} \int_0^T \theta_s^2 ds + \int_0^T \theta_s dB_s}.
\eew

Thus we have
\bew
E_{Q_k} [ \Phi(S_T)] = \sup_{{Q_\theta} \in \mathcal P}  E_{Q_\theta} [ \Phi(S_T)]:=\mathcal {\bar E}
[\Phi(S_T)],
\eew since $Q_k \in \mathcal P$ and $|\theta_t|\leq k$.
In the similar fashion, we can also show that 
\bew
E_{Q_{-k}} [ \Phi(S_T)] = \inf_{{Q} \in \mathcal P}  E_{Q} [ \Phi(S_T)]:=\mathcal {\underline E}
[\Phi(S_T)],
\eew where $Q_{-k}$ is defined as
\bew
\f{d Q_{-k}}{dP} = e^{-\f{1}{2} k^2 T -  k B_T}.
\eew
Now suppose that $\Phi$ is a decreasing function. 
Then $-\Phi$ is an increasing function. 
So we have
\bew
&&\mathcal {\bar E}[\Phi(S_T)] =-\mathcal {\underline E}
[-\Phi(S_T)]  = -E_{Q_{-k}} [ -\Phi(S_T)] = E_{Q_{-k}} [ \Phi(S_T)],\\
&&\mathcal {\underline E}[\Phi(S_T)] = -\mathcal {\bar E}[-\Phi(S_T)]
=-E_{Q_{k}} [ -\Phi(S_T)]=E_{Q_{k}} [ \Phi(S_T)].
\eew

\vspace{-2mm}
\section*{Acknowledgment}
\vspace{-3mm}
This work was supported by the research grant of
          Sungshin Women's University in 2018.




\begin{thebibliography}{10}












\bibitem{ChenChenDavison05}
Z. Chen, T. Chen and M. Davison, Choquet expectation and Peng's $g$-expectation,
The Annals of Probability 33 (2005) 1179--1199.



\bibitem{CerdaMartinSilvestre11}
J. Cerda, J. Martin \& P. Silvestre,  Capacitary function spaces, Collect. Math. 62 (2011) 95-118. https://doi.org/10.1007/s13348-010-0031-7



\bibitem{Choquet53}
G. Choquet, Theory of capacities, Ann. Inst. Fourier (Grenoble) 5 (1953) 131--195.

\bibitem{CoquetHuMeminPeng02}
F. Coquet, Y. Hu, J. M\'{e}min and S. Peng, Filtration consistent nonlinear expectations and related $g$-expectations, Probability Theory and Related Fields 123 (2002) 1--27.







\bibitem{FollmerSchied04}
H. F\"{o}llmer and A. Schied, Stochastic Finance, An introduction in discrete time, Walter de Gruyter, Berlin, 2004.





\bibitem{RosazzaGianin06}
E. R. Gianin, Some examples of risk measures via $g$-expectations,
 Insurance: Mathematics and Economics 39 (2006) 19--34.


\bibitem{HeHuChen09}
K. He, M. Hu and Z. Chen, The relationship between risk measures and Choquet expectations in the 
framework of $g$-expectations, Statistics and Probability Letters 79 (2009) 508--512.





\bibitem{Jiang06}
L. Jiang, Convexity, translation invariance and subadditivity for $g$-expectation and  related risk measures, Annals of Applied Provability 18 (2006) 245--258.
https://arxiv.org/abs/0801.3340
 

\bibitem{LongJiang09}
L. Jiang, A necessary and sufficient condition for probability measures dominated by $g$-expectation, Statistics \& Probability Letters 79 (2009) 196--201. 
https://doi.org/10.1016/j.spl.2008.07.037


\bibitem{KarouiPengQuenez97}
N. El Karoui, S. Peng, M.C. Quenez,  Backward Stochastic Differential Equations in Finance, Mathematical Finance  7 (1997) 1--71.
 https://doi.org/10.1111/1467-9965.00022




\bibitem{HongxiaWang15}
H. Wang, Conditional Choquet Expectation, Communications in Statistics - Theory and Methods 44 (2015) 3782--3795. 
DOI: 10.1080/03610926.2014.935432 


\bibitem{ChenKulperger06}
Z. Chen \& R. Kulperger, Minimax pricing and Choquet pricing, 
Insurance: Mathematics and Economics 38 (2006) 518--528.
https://doi.org/10.1016/j.insmatheco.2005.11.010.


\bibitem{PardouxPeng90}
E. Pardoux, \& S. Peng, Adapted solution of a backward stochastic differential equation,
 Systems and Control Letters 14 (1990) 55--61.

\bibitem{Peng97}
S. Peng, Backward SDE and related g-expectation, backward stochastic DEs,
Pitman 364 (1997) 141--159.


\bibitem{Schmeidler86}
D. Schmeidler, Integral Representation without Additivity,
 Proceedings of the American Mathematical Society 97 (1986) 255--261.












\end{thebibliography}


\bibliographystyle{amsplain}

\end{document}